\documentclass[twoside,twocolumn,9pt]{article}
\usepackage[english]{babel} % This should come before natmove
\usepackage{extsizes}
\usepackage[super,sort&compress,comma]{natbib}
\usepackage{natmove} % move superscript citations after punctuation
\usepackage[version=3]{mhchem}
\usepackage[left=1.5cm, right=1.5cm, top=1.785cm, bottom=2.0cm]{geometry}
\usepackage{balance}
\usepackage{mathptmx}
\usepackage{sectsty}
\usepackage{graphicx} 
\usepackage{lastpage}
\usepackage[format=plain,justification=justified,singlelinecheck=false,font={stretch=1.125,small,sf},labelfont=bf,labelsep=space]{caption}
\usepackage{float}
\usepackage{fancyhdr}
\usepackage{fnpos}
\addto{\captionsenglish}{%
  
}
\usepackage{array}
\usepackage{droidsans}
\usepackage{charter}
\usepackage[T1]{fontenc}
\usepackage[usenames,dvipsnames]{xcolor}
\usepackage{setspace}
\usepackage[compact]{titlesec}
\usepackage{hyperref}
\usepackage{amssymb}
\usepackage{amsmath}
\usepackage{verbatim}
\usepackage{bm}
\usepackage{siunitx}
\usepackage{listings}
\usepackage{hyperref}
\hypersetup{
   colorlinks=true,
   urlcolor=blue,    % color of external links
   linkcolor=black,  % color of toc, list of figs etc.
   citecolor=blue,   % color of links to bibliography
}

\usepackage{epstopdf}%This line makes .eps figures into .pdf - please comment out if not required.

\definecolor{cream}{RGB}{222,217,201}

\begin{document}

\pagestyle{fancy}
\thispagestyle{plain}
\fancypagestyle{plain}{
%%%HEADER%%%
\renewcommand{\headrulewidth}{0pt}
}
%%%END OF HEADER%%%

%%%PAGE SETUP - Please do not change any commands within this section%%%
\makeFNbottom
\makeatletter
\renewcommand\LARGE{\@setfontsize\LARGE{15pt}{17}}
\renewcommand\Large{\@setfontsize\Large{12pt}{14}}
\renewcommand\large{\@setfontsize\large{10pt}{12}}
\renewcommand\footnotesize{\@setfontsize\footnotesize{7pt}{10}}
\makeatother

\renewcommand{\thefootnote}{\fnsymbol{footnote}}
\renewcommand\footnoterule{\vspace*{1pt}% 
\color{cream}\hrule width 3.5in height 0.4pt \color{black}\vspace*{5pt}} 
\setcounter{secnumdepth}{5}

\makeatletter 
\renewcommand\@biblabel[1]{#1}            
\renewcommand\@makefntext[1]% 
{\noindent\makebox[0pt][r]{\@thefnmark\,}#1}
\makeatother 
\renewcommand{\figurename}{\small{Fig.}~}
\sectionfont{\sffamily\Large}
\subsectionfont{\normalsize}
\subsubsectionfont{\bf}
\setstretch{1.125} %In particular, please do not alter this line.
\setlength{\skip\footins}{0.8cm}
\setlength{\footnotesep}{0.25cm}
\setlength{\jot}{10pt}
\titlespacing*{\section}{0pt}{4pt}{4pt}
\titlespacing*{\subsection}{0pt}{15pt}{1pt}
%%%END OF PAGE SETUP%%%

%%%FOOTER%%%
\fancyfoot{}
\fancyfoot[LO,RE]{\vspace{-7.1pt}\includegraphics[height=9pt]{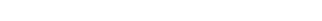}}
\fancyfoot[CO]{\vspace{-7.1pt}\hspace{13.2cm}\includegraphics{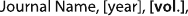}}
\fancyfoot[CE]{\vspace{-7.2pt}\hspace{-14.2cm}\includegraphics{head_foot/RF}}
\fancyfoot[RO]{\footnotesize{\sffamily{1--\pageref{LastPage} ~\textbar  \hspace{2pt}\thepage}}}
\fancyfoot[LE]{\footnotesize{\sffamily{\thepage~\textbar\hspace{3.45cm} 1--\pageref{LastPage}}}}
\fancyhead{}
\renewcommand{\headrulewidth}{0pt} 
\renewcommand{\footrulewidth}{0pt}
\setlength{\arrayrulewidth}{1pt}
\setlength{\columnsep}{6.5mm}
\setlength\bibsep{1pt}
%%%END OF FOOTER%%%

%%%FIGURE SETUP - please do not change any commands within this section%%%
\makeatletter 
\newlength{\figrulesep} 
\setlength{\figrulesep}{0.5\textfloatsep} 

\newcommand{\topfigrule}{\vspace*{-1pt}% 
\noindent{\color{cream}\rule[-\figrulesep]{\columnwidth}{1.5pt}} }

\newcommand{\botfigrule}{\vspace*{-2pt}% 
\noindent{\color{cream}\rule[\figrulesep]{\columnwidth}{1.5pt}} }

\newcommand{\dblfigrule}{\vspace*{-1pt}% 
\noindent{\color{cream}\rule[-\figrulesep]{\textwidth}{1.5pt}} }

\makeatother
%%%END OF FIGURE SETUP%%%

%%%TITLE, AUTHORS AND ABSTRACT%%%
\twocolumn[
  \begin{@twocolumnfalse}
{\includegraphics[height=30pt]{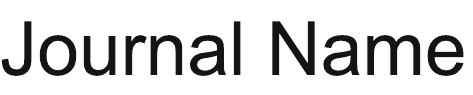}\hfill\raisebox{0pt}[0pt][0pt]{\includegraphics[height=55pt]{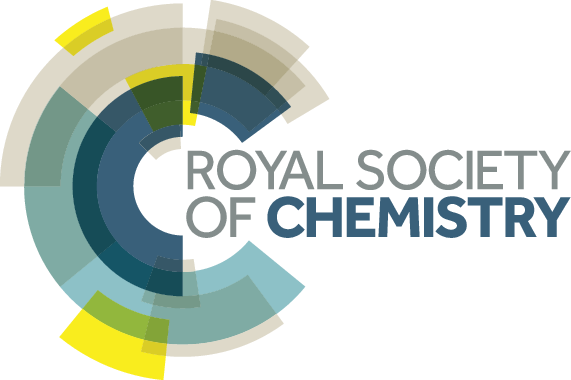}}\\[1ex]
\includegraphics[width=18.5cm]{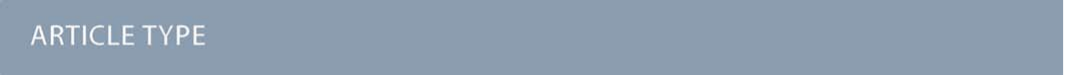}}\par
\vspace{1em}
\sffamily
\begin{tabular}{m{4.5cm} p{13.5cm} }

\includegraphics{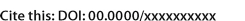} & \noindent\LARGE{\textbf{Teacher-student training improves accuracy and efficiency of machine learning interatomic potentials$^\dag$}} \\%Article title goes here instead of the text "This is the title"
\vspace{0.3cm} & \vspace{0.3cm} \\

 & \noindent\large{
    Sakib Matin$^{\ast}$\textit{$^{a}$}, 
    Alice E. A. Allen\textit{$^{a,b,c}$}, 
    Emily Shinkle\textit{$^{d}$}, 
    Aleksandra Pachalieva\textit{$^{e}$},
    Galen T. Craven\textit{$^{a}$}, 
    Benjamin Nebgen\textit{$^{a}$}, 
    Justin S. Smith\textit{$^{f}$}, 
    Richard Messerly\textit{$^{a}$}, 
    Ying Wai Li\textit{$^{d}$}, 
    Sergei Tretiak\textit{$^{a,b,g}$}, 
    Kipton Barros\textit{$^{a,b}$}, 
    and Nicholas Lubbers\textit{$^{d}$}
} \\

%%% ABSTRACT %%%
\includegraphics{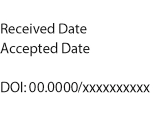} & \noindent\normalsize{
Machine learning interatomic potentials (MLIPs) are revolutionizing the field of molecular dynamics (MD) simulations. Recent MLIPs have tended towards more complex architectures trained on larger datasets. The resulting increase in computational and memory costs may prohibit the application of these MLIPs to perform large-scale MD simulations. Here, we present a teacher-student training framework in which the latent knowledge from the teacher (atomic energies) is used to augment the students' training. We show that the light-weight student MLIPs have faster MD speeds at a fraction of the memory footprint compared to the teacher models. Remarkably, the student models can even surpass the accuracy of the teachers, even though both are trained on the same quantum chemistry dataset. Our work highlights a practical method for MLIPs to reduce the resources required for large-scale MD simulations.
} \\
\end{tabular}

 \end{@twocolumnfalse} \vspace{0.6cm}

  ]
%%%END OF TITLE, AUTHORS AND ABSTRACT%%%

%%%FONT SETUP - please do not change any commands within this section
\renewcommand*\rmdefault{bch}\normalfont\upshape
\rmfamily
\section*{}
\vspace{-1cm}

%%%FOOTNOTES%%%
\footnotetext{\textit{$^{a}$~Theoretical Division, Los Alamos National Laboratory, Los Alamos, New Mexico, USA.}}
\footnotetext{\textit{$^{b}$~Center for Nonlinear Studies, Los Alamos National Laboratory, Los Alamos, New Mexico 87546}}
\footnotetext{\textit{$^{c}$ Max Planck Institute for Polymer Research, Ackermannweg 10, 55128 Mainz}}

\footnotetext{\textit{$^{d}$~Computer, Computational, and Statistical Sciences Division, Los Alamos National Laboratory, Los Alamos, New Mexico, USA.}}
\footnotetext{\textit{$^{e}$~Earth and Environmental Sciences Division, Los Alamos National Laboratory, Los Alamos, New Mexico, USA.}}
\footnotetext{\textit{$^{f}$~Nvidia Corporation, Santa Clara, California, 9505, USA}}
\footnotetext{\textit{$^{g}$~Center for Integrated Nanotechnologies, Los Alamos National Laboratory, Los Alamos, New Mexico 87546}}

%\footnotetext{\textit{$^{a}$~Address, Address, Town, Country. Fax: XX XXXX XXXX; Tel: XX XXXX XXXX; E-mail: xxxx@aaa.bbb.ccc}}
%\footnotetext{\textit{$^{b}$~Address, Address, Town, Country. }}

%Please use \dag to cite the ESI in the main text of the article.
%If you article does not have ESI please remove the the \dag symbol from the title and the footnotetext below.
%\footnotetext{\dag~Supplementary Information available: [details of any supplementary information available should be included here]. See DOI: 00.0000/00000000.}
%additional addresses can be cited as above using the lower-case letters, c, d, e... If all authors are from the same address, no letter is required

\footnotetext{${\ast}$~Correspondence should be addressed to sakibmatin@gmail.com}
%%%END OF FOOTNOTES%%%

%%%MAIN TEXT%%%%
% ================================================================= %
\section{Introduction\label{sec:intro}}
Molecular Dynamics (MD)~\cite{allen2017computer} is ubiquitous in chemistry~\cite{tuckerman2000understanding}, materials science~\cite{steinhauser2009review}, and drug discovery~\cite{de2016role} as well as other fields. 
Accurate chemical and thermodynamic properties derived from MD rely on accurate interatomic potentials, which parameterize the many-body interactions present between atoms~\cite{allen2017computer,thompson2022lammps}. 
Simulation scales may vary greatly depending on the questions of interest and available resources, and there is a persistent need for greater model efficiency. Traditional classical potentials are very fast, making it possible to perform large-scale simulations of billions of atom~\cite{jung2019scaling}, or to perform hundreds of millions of integration time-steps for small  systems.
Recently, there is great demand for interatomic potentials that are more accurate via machine learning models that are trained to reference quantum mechanical force calculations. Here, the goals of efficiency and accuracy can be in conflict.
Our paper is concerned with techniques for improving the efficiency of the interatomic potential {\it without} sacrificing accuracy. 

The gold-standard for computational chemistry is 
\textit{ab initio} molecular dynamics, 
which
uses quantum chemistry (QC) methods for accurately calculating interatomic forces from first principles~\cite{szabo2012modern, burke2012perspective}. However, the often prohibitive computational cost of QC methods, and its rapid growth with system size, limits the size of the systems that can be simulated. Machine learning interatomic potentials (MLIPs)~\cite{zuo2020performance, deringer2021gaussian, fedik2022extending, kulichenko2021rise, unke2021machine} can be trained on QC datasets to map from an atomic configuration to energy and forces.
MLIPs can achieve the chemical accuracy (error $< 1 \mathrm{kcal/mol}$)  of QC simulations~\cite{chigaev2023lightweight}
at drastically reduced computational costs. Most MLIPs achieve linear scaling with system size by utilizing the approximation that the total predicted energy can be decomposed as a sum of spatially local atom-wise or pair-wise contributions~\cite{behler2016perspective, kulichenko2021rise,fedik2022extending, ibayashi2023allegro}. 
MLIPs have seen explosive growth and have been successfully applied to predicting potential energy surfaces~\cite{behler2007generalized, bartok2010gaussian, rupp2012fast, smith2017ani, schutt2018schnet, lubbers2018hierarchical, smith2021automated, kovacs2021linear, batzner2022E3-equivariant}, with extensions to a variety of quantities, such as charges~\cite{unke2019physnet, yao2018tensormol, sifain2018discovering, ko2021fourth, ko2023accurate}, and more~\cite{eckhoff2020predicting,rezajooei2023neural,rackers2023recipe,shinkle2024thermodynamic,magedov2021bond}. However, MLIPs only mitigate the cost of QC---they do not eliminate it. Generating training data for MLIPs is costly due to the steep scaling of QC methods in system size $N$, e.g., $\mathcal{O}(N^7)$ at the CCSD(T) level of theory~\cite{szabo2012modern}. Therefore any approach which uses data more efficiently can potentially reduce the overall computational costs of MLIPs. 

Furthermore, a recent trend in the field is the significant effort to construct foundation model MLIPs~\cite{batatia2023foundation, allen2024learning, zhang2024pretraining} with broad chemical coverage\cite{jain2013commentary}, inspired by the success of large pre-trained models in natural language processing~\cite{bommasani2021opportunities}. 
Foundation MLIPs~\cite{batatia2022mace}, which may be parameterized by up to $10^9$ fitting parameters~\cite{sriram2022towards}, have high computational and memory requirements~\cite{kelvinius2023accelerating, amin2025towards}. These increasing computational and memory costs compared to classical force fields~\cite{allen2017computer} can limit the usability of MLIPs for large scale MD simulations~\cite{unke2021machine, xie2023ultra, kelvinius2023accelerating, amin2025towards}. 

\begin{figure*}[h]
    \centering
    \includegraphics[width=1.0\textwidth]{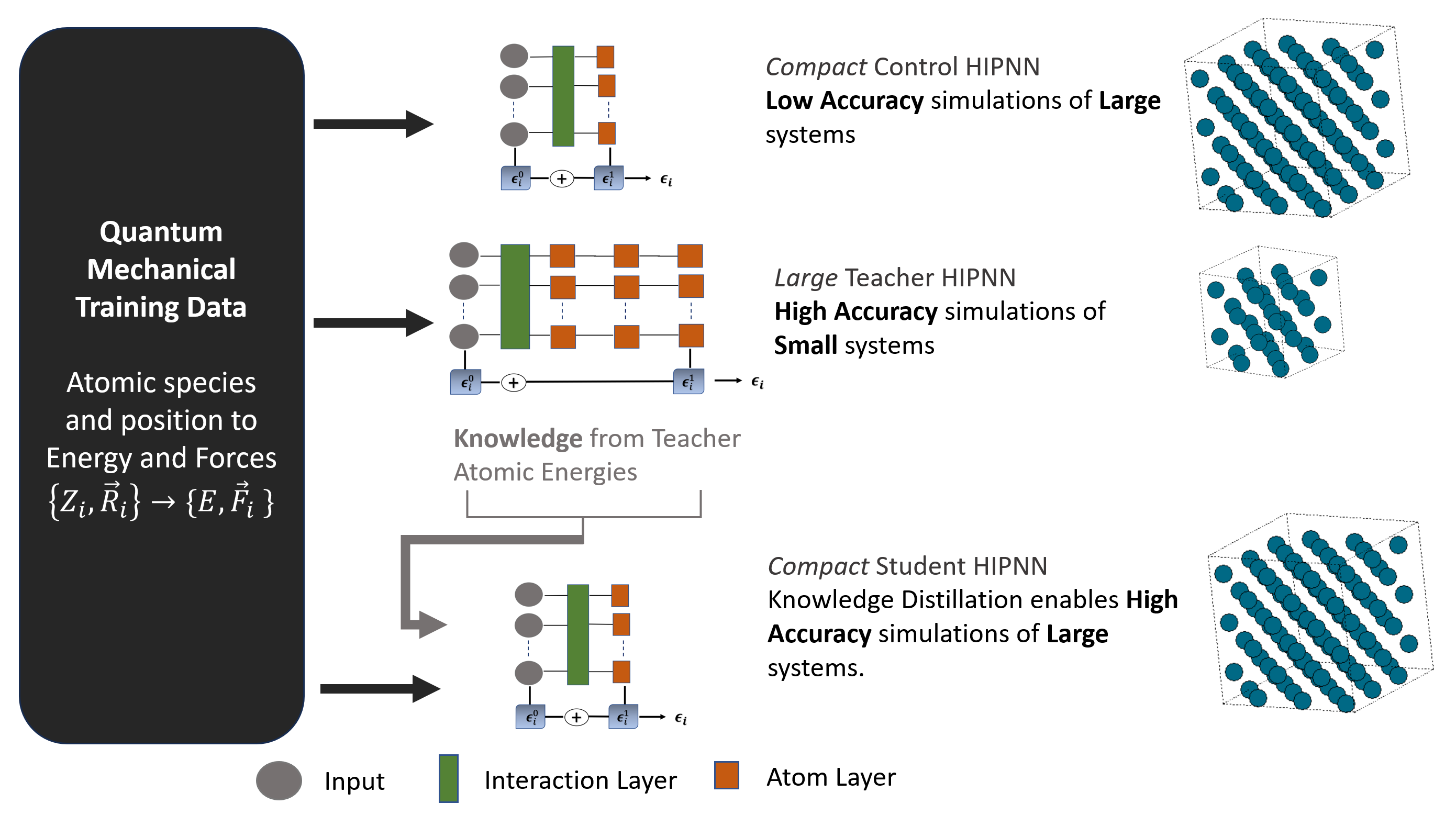}
    \caption{\textit{Teacher-student training for Hierarchically Interacting Particle Neural Network (HIPNN) machine learning interatomic potential.} A HIPNN model consists of an input node, a message passing interaction layer and feed-forward regression layers called atom layers. The teacher HIPNN is trained on the quantum mechanical energy and forces data, and generates latent knowledge in the form of atomic energies. This augments the student model training to improve the accuracy. The student HIPNN that has fewer trainable weights, which contributes to faster inference and lower memory requirements. The control HIPNNs have the same number of trainable parameters as the student but are trained only on ground truth data. We show that the student models are more accurate than the control models. 
    }
    \label{fig:overview}
\end{figure*}

In this manuscript, we introduce a teacher-student training method for MLIP with a central purpose of building the MLIPs with faster inference and lower memory requirements. The teacher-student training is a class of methods~\cite{hinton2015distilling, furlanello2018born, yang2020distilling} to reduce the inference costs of different ML models and more effectively utilize existing datasets. 
An initial teacher model is trained and then used to augment the training of a student model that may have faster inference~\cite{yang2020distilling,kelvinius2023accelerating, amin2025towards}, smaller memory requirements~\cite{sanh2019distilbert}, or better generalization capacity~\cite{furlanello2018born}. 
Crucially, this knowledge distillation procedure does not require additional first-principles training data. Instead, auxiliary predictions of the teacher model are used to augment the data used for training of the student model. The innovation in this work is to use the teacher's local atomic energy predictions as the auxiliary training data. Although these local atomic energies have traditionally been considered a latent feature of the MLIP more~\cite{fedik2022extending}, the present work highlights that they carry important information.
Note that the latent atomic energies provided to the student are far greater in number than the single global QC energy. Thus the student is trained using a significantly larger number of constraints than the teacher. Because the student model will typically have fewer trainable weighs than the teacher model, this approach can yield significant gains for inference speed and memory requirements.
We also find that the student model can achieve higher-accuracy than the teacher. An overview of our workflow is represented in Fig~\ref{fig:overview}.

% ================================================================= %
\section{Methods \& Background \label{sec:methods}}
We apply the teacher-student training to an MLIP architecture, namely the Hierarchically Interacting Particle Neural Network (HIPNN)~\cite{lubbers2018hierarchical, chigaev2023lightweight}. HIPNN is a message passing graph neural network used to model atomistic systems~\cite{lubbers2018hierarchical,sifain2018discovering}, and recent variants incorporate equivariant tensor sensitivity for higher accuracy~\cite{chigaev2023lightweight}. The architecture of HIPNN is briefly reviewed in in Sec.~\ref{sec:architecture_hipnn}. In Sec.~\ref{sec:teacher-student}, we introduce the teacher-student paradigm. Section~\ref{sec:training_and_loss} summarizes the training procedure. Sec.~\ref{sec:data} details the datasets studied. Section~\ref{sec:results} contains a systematic exploration of the teacher-student training for the HIPNN model.  

\subsection{Architecture\label{sec:architecture_hipnn}}
HIPNN~\cite{lubbers2018hierarchical} is a message-passing graph neural network~\cite{duval2023hitchhiker} that can map atomic configurations to various chemical properties such as energy~\cite{lubbers2018hierarchical}, forces~\cite{chigaev2023lightweight}, dipoles~\cite{sifain2018discovering} etc. Similar to many MLIP architectures, the interaction layers of HIPNN allow for mixing of atomic environments between neighbors via message passing to construct the learnable features~\cite{lubbers2018hierarchical}. 
The local atomic environment of each atom is initially featurized using atomic number (in HIPNN, as a one-hot encoding, although also common is a random embedding) and passed through several layers along with pair-wise displacement vectors between the neighbors with local cut-off to predict the atom energy $\mathcal{\epsilon}_i$ for each atom $i$.
Automatic differentiation is used to calculate the forces on each atom from the total molecular energy. 

MLIP architectures almost universally infer a local decomposition in their predictions of extensive quantities. In message passing neural network parlance, the readout function is a linear summation over node states. Specifically, in HIPNN, the energy $E[\bm{r}_1, \bm{r}_2, \ldots, \bm{r}_N]$ of a configuration, where $N$ is the number of atoms, is decomposed into a sum over local contributions,
\begin{align}
    E \approx \hat{E} = \sum^{N}_i \mathcal{\epsilon}_i .
\end{align}
The energy contributions are formed by linear combination of the features (also called embeddings) learned by the neural network. The teacher-student procedure we investigate here depends on the existence of {\it some} local energy decomposition, but is flexible about its details. For example, our procedure could also work on models such as Allegro that decompose energy into contributions on local bonds~\cite{ibayashi2023allegro}.

In the original publication~\cite{lubbers2018hierarchical}, HIPNN only used scalar pair-wise distances between neighbors, which captures a subset of the higher-order many-body information contained in the local chemical environment~\cite{batatia2022mace, chigaev2023lightweight}. Subsequently, HIPNN with Tensor Sensitivity~\cite{chigaev2023lightweight} utilizes higher order tensor products of the displacement vectors between atoms to construct more informative descriptors.  The hyper-parameter $l_\mathrm{max}$ corresponds to the highest order tensor used in HIPNN. A weight tying scheme between the different order sensitivity functions leads to only modest increase in number of trainable parameters, thereby reducing the training and evaluation costs, and improving data efficiency~\cite{chigaev2023lightweight}. The $l_\mathrm{max}=0$ HIPNN model utilizes only scalar pair-wise distances and coincides with the model developed in original publication~\cite{lubbers2018hierarchical}. In this work we select $l_\mathrm{max}=1$ to allow for vector sensitivity. Other hyper-parameters for the HIPNN models used in this paper are given in Sec.~\ref{app:hyper}.

% 2 column formatting requirements
\begin{figure*}[h]
    \includegraphics[]{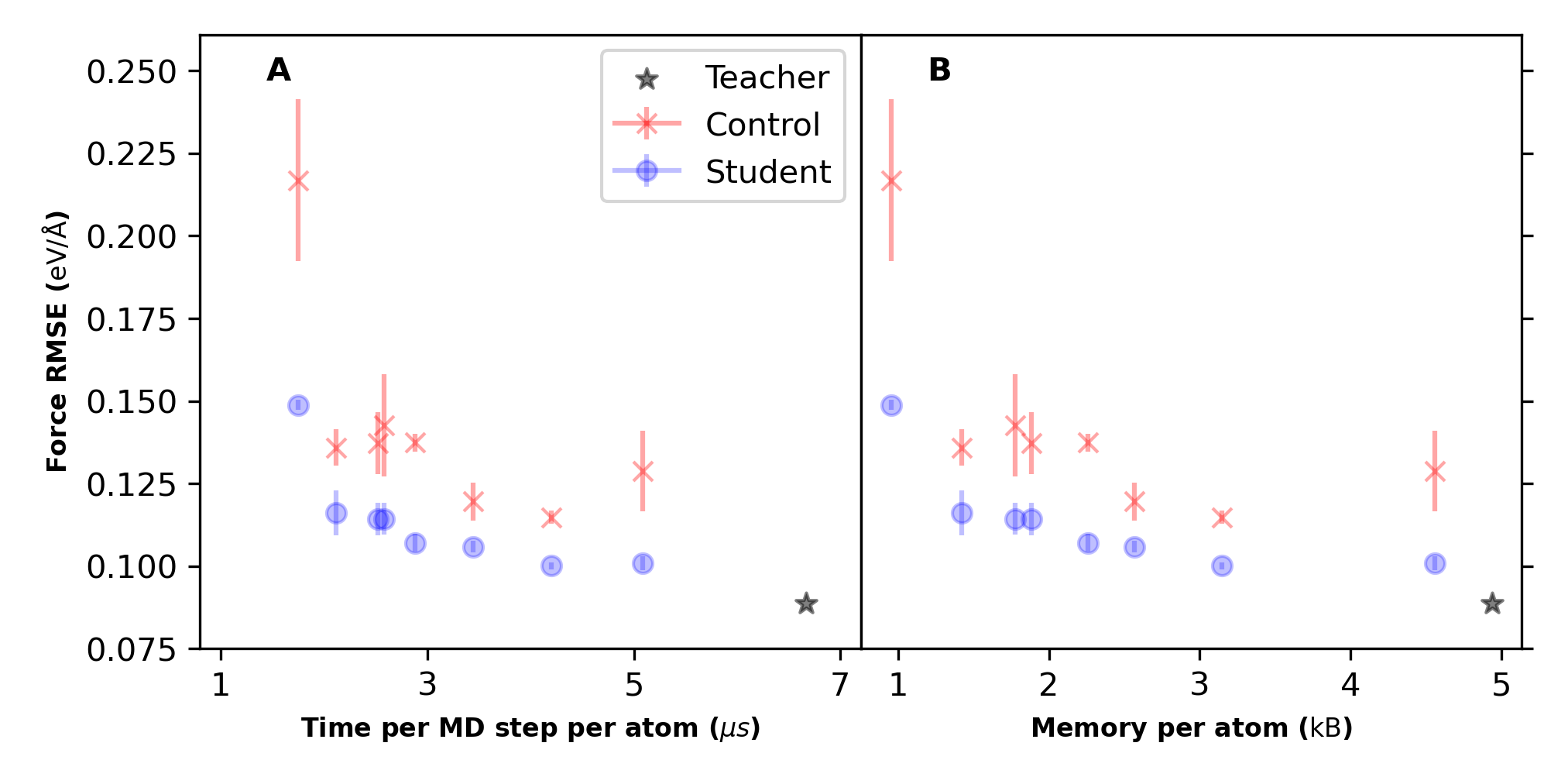}
\caption{
\textit{The student MLIPs are Pareto dominant compared to the control models.} The student and control MLIPs are trained on the ANI-Al aluminum dataset~\cite{smith2021automated}. The accuracy metric, force RMSE, is plotted as a function of the efficiency metrics, namely, time per MD step per atom in (A) and memory per atom in (B). The origin of the plots corresponds to the Pareto optimum solution. The MD simulations are run using the ASE~\cite{larsen2017atomic} code on a  Nvidia A6000 (48GB) GPU. The error bar corresponds to the standard deviation of an ensemble of $4$ models, which differ by random weight initializations. }
\label{fig_pareto}
\end{figure*}

% ================================================================= %
\subsection{Teacher-Student Training\label{sec:teacher-student}}
In the teacher-student training method, a pre-trained teacher model (or models) is used to train a student model to improve speed~\cite{yang2020distilling,kelvinius2023accelerating, amin2025towards}, memory requirements~\cite{sanh2019distilbert}, and generalization~\cite{furlanello2018born}. Knowledge Distillation (KD)~\cite{hinton2015distilling} is a well known form of teacher-student training. In the original KD publication~\cite{hinton2015distilling} auxiliary targets are the teacher model's predicted relative class probabilities (before the application of the softmax operation in the output layer). The auxiliary targets generated by the teacher contain richer information about the structure of the classes in the dataset. The student models were trained on both the ground truth data and the auxiliary targets and performed better than the control models, which have same architecture as the student but are trained only on the ground truth data. Later works have incorporated deeper architecture dependent knowledge~\cite{ sanh2019distilbert, gou2020knowledge, yang2020distilling}. KD has been successfully applied to many different architectures such as CNN~\cite{xu2020feature},  graph neural networks~\cite{yang2020distilling}, transformers~\cite{sanh2019distilbert} and more~\cite{gou2021knowledge}. 
% The teacher-student training framework has been used by the the ORCA large language model learn to mimic the GPT-4 model's reasoning via explanation traces~\cite{mukherjee2023orca}. 
Another variant of the teacher-student training is the ``Born Again'' (BA) method~\cite{furlanello2018born},  where the student and teacher models have the same architecture, and the student model can surpass the teacher's accuracy. Multiple teachers can be used to train a single student model to improve performance~\cite{chebotar2016distilling, furlanello2018born, gong2025predictive, matin2025ensemble}. Previous works focus on classification tasks on images~\cite{hinton2015distilling} and text~\cite{sanh2019distilbert}, and limit applications to regression tasks on graph structured data~\cite{xu2022contrastive}. 

Recently, teacher-student methods have been explored in chemistry for accelerating molecular dynamics~\cite{kelvinius2023accelerating, amin2025towards}, physics-constrained data augmentation~\cite{f2025improving}, and material property prediction~\cite{zhu2024addressing}. In a related work in Ref.~\citenum{gardner2023synthetic}, the teacher MLIP (trained on QC ground truth) is used to generate synthetic data by running MD under different conditions. The student model is initially pre-trained on the synthetic data then fine-tuned on the QC ground truth. 

In the first step of our teacher-student method, we train a complex ``teacher'' MLIP on a QC dataset, which consists of configuration energy and forces on each atom. This teacher MLIP can generate auxiliary targets, namely per-atom energies. The atomic energy may provide more fine-grained information than the aggregate configuration energy~\cite{kelvinius2023accelerating, gardner2023synthetic, jung2024atomic}. Then, we train the student MLIP on the original QC data and auxiliary targets generated by the teacher MLIP. 
This economical approach does not require any expensive QC calculations beyond the original dataset needed to train the teacher model, nor exhaustive hyper-parameter tuning.

% ================================================================= %
\subsection{Training Procedure and Loss Function\label{sec:training_and_loss}}
We train the teacher model $\mathcal{T}$ on the QC dataset $\mathfrak{D}$, which contains the set of atomic position and species for each configuration and the corresponding energy and the forces per atom, $\mathfrak{D}:\left\{\bm{R}_i, Z_i\right\} \to \left\{E, \bm{F}_i\right\}$.

The NN model is trained in a standard manner using stochastic gradient descent on the loss function
\begin{align}
    \mathcal{L}_\mathrm{Teacher} &= w_E \mathcal{L}_\mathrm{err}(\hat{E},E)  + w_F \mathcal{L}_\mathrm{err}(\hat{F},F)  + w_{L_2}\mathcal{L}_{L_2} + w_R \mathcal{L}_{R}. 
    \label{eq_loss}
\end{align}
The $\mathcal{L}_{L_2}$ loss term is a regularization of the model weights, which is commonly added to loss functions to reduce over-fitting. The $\mathcal{L}_\mathrm{R}$ term, specific to HIPNN, also enhances model stability. The error loss $\mathcal{L}_\mathrm{err}$ could be any combination of common metrics, but in our case is an equal weighting between root-mean-squared error (RMSE) and mean-absolute error (MAE) losses,
\begin{align}
\mathcal{L}_\mathrm{err}(\hat{V},V) = \mathrm{RMSE}(\hat{V},V)+ \mathrm{MAE}(\hat{V},V),
\end{align}
where $\hat{V}$ is the model prediction and $V$ is the ground truth target. Our loss function uses a sum of the RMSE and MAE errors, similar to the original HIPNN model publications~\cite{lubbers2018hierarchical, chigaev2023lightweight}. The weights of each term in the loss are listed in Appendix~\ref{app:hyper}. 

The teacher $\mathcal{T}$ maps the local atomic environments to atomic energies $\epsilon_i$ which are summed over to obtain the energy $\hat{E}$. Although $\epsilon_i$ is not directly a physical observable, it nonetheless captures important information about how the local geometry affects the final prediction of the model~\cite{kelvinius2023accelerating, gardner2023synthetic, jung2024atomic}. Here, we use the teacher MLIP's atomic energies as the knowledge to be transferred to the student MLIPs. 

The atom energy $\epsilon_i$ predictions from $\mathcal{T}$ are used to construct the augmented dataset
\begin{align}
     \mathfrak{\tilde{D}}: \left\{\bm{R}_i, Z_i\right\} \to \left\{E, \epsilon^\mathcal{T}_i, \bm{F}_i\right\}.
\end{align}
Note that the original dataset $\mathfrak{D}$ has the same set of configurations (inputs) as the augmented data $\mathfrak{\tilde{D}}$. 

We train the student models $\mathcal{S}$ on the augmented dataset $\mathfrak{\tilde{D}}$ with the loss function

\begin{align}
    \mathcal{L}_\mathrm{Student} &= w_E \mathcal{L}_\mathrm{err}(\hat{E},E)  + w_F \mathcal{L}_\mathrm{err}(\hat{F},F) + w_A \mathcal{L}_\mathrm{err}(\epsilon^\mathcal{S},\epsilon^\mathcal{T})
    \nonumber \\ 
    & \quad + w_{L_2}\mathcal{L}_{L_2} + w_R \mathcal{L}_{R} 
\end{align}
This captures the notion that the student should learn from the teacher using the loss term $\mathcal{L}_\mathrm{err}(\epsilon^\mathcal{S},\epsilon^\mathcal{T})$ which encourages the partitioning of energy among atomic sites in the student, $\epsilon^\mathcal{S}$, to match that of the teacher, $\epsilon^\mathcal{T}$.
We explore student models by varying several architectural parameters of the NN, namely, the layer width $n_\mathrm{feature}$, the number of sensitivities function $n_d$ used for the distance embedding, and the number of atom layers used after interaction layers, $n_\mathrm{atom\_layer}$.
To assess the effectiveness of the method, we also train control models $\mathcal{C}$ that have the same architecture as the students $\mathcal{S}$ but are trained only on the dataset $\mathfrak{D}$ using $\mathcal{L}_\mathcal{T}$; everything else is held constant in these models except for the student-teacher loss. The accuracy of the control models serves as a benchmark to compare the efficacy of the teacher-student training framework.

% ================================================================= %
\subsection{Data\label{sec:data}}
We apply our teacher-student workflow on the ANI-Al dataset~\cite{smith2021automated}, which consists of condensed-phase aluminum geometries, with energies and forces calculated using Density Functional Theory (DFT). 
% Smith et. al. 
The authors of reference~\cite{smith2021automated} created the dataset 
using an automated active learning framework to generate adequate coverage of the configurational space. In this workflow, an ensemble of ANI  models~\cite{smith2017ani} were trained to the initial DFT dataset, which consisted of random structures.
The main loop of the active learning workflow involves running MD simulations with the ensemble under varying thermodynamic conditions (time-dependent temperatures and density schedules) on boxes of $\approx50$ to $\approx 250$ atoms. New DFT calculations were performed on configurations where the ensemble disagreement exceeded a predefined threshold. Then the MLIPs were retrained to the expanded dataset. This loop terminates when long MD simulations ($250 \mathrm{ps}$) could be performed without identifying any new configurations with high ensemble disagreement. 
Over 50 generations of models, the final dataset is comprised of about 6,000 DFT calculations with the Perdew-Burke-Ernzerhof functional. The dataset is available online~\cite{atomistic-ml2021ani-al}. More details are available in the original publication describing the construction of the dataset.~\cite{smith2021automated}

\begin{figure}[]
    \includegraphics[]{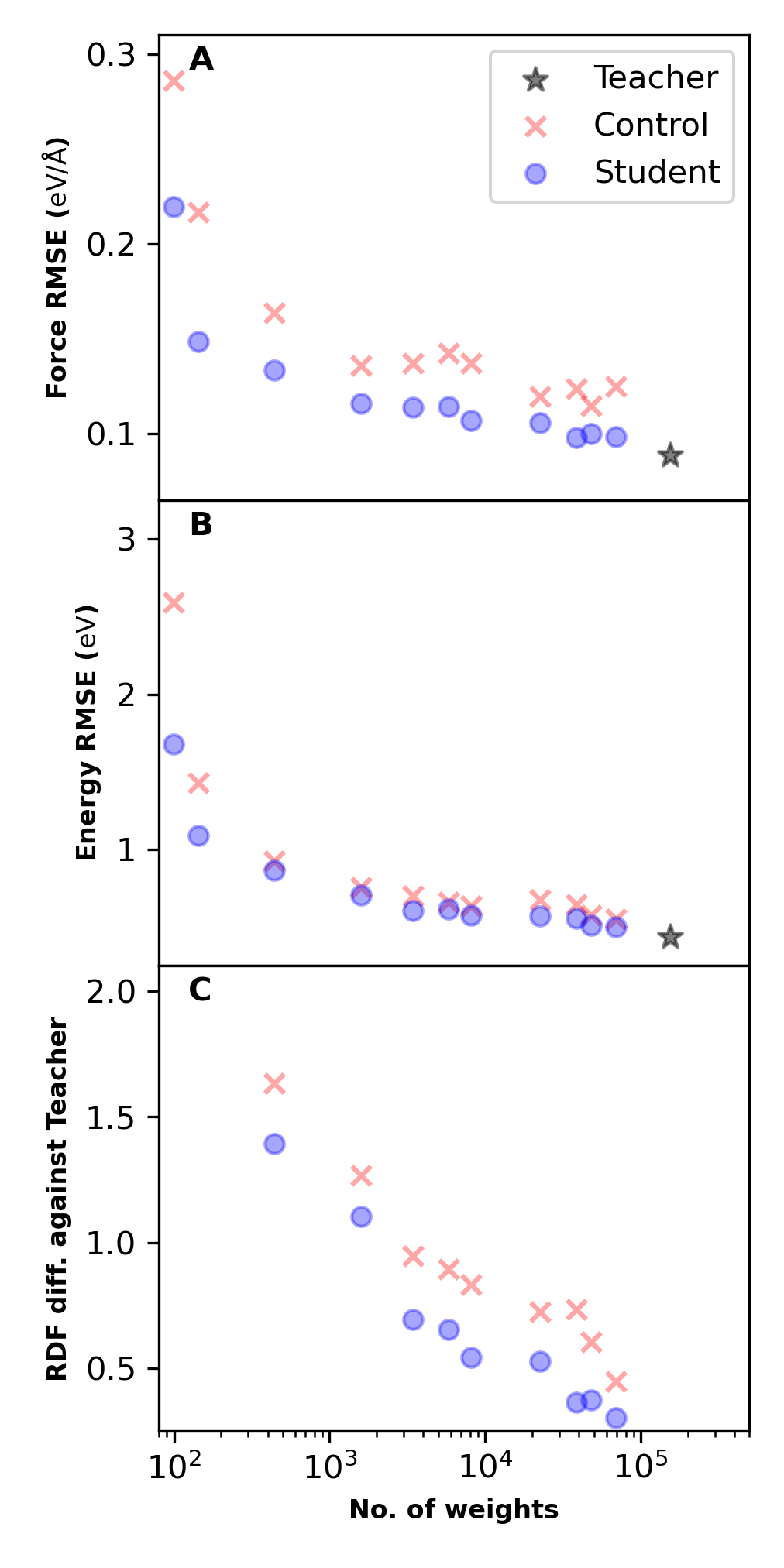}
\caption{\textit{The student MLIPs are more accurate than the control architectures at the corresponding capacity}. The number of trainable weights captures the model capacity. The force RMSE errors and energy per-atom RMSE errors for the training dataset are plotted in panels (A) and (B) respectively. The total absolute error of the radial distribution function (RDF) of the student and control MLIPs compared to the teacher MLIP is shown in panel (C). The error metrics with respect to the training datasets in (A) and (B) show similar trends as the MD-based accuracy metrics in (C). 
}
\label{fig_accuracy}
\end{figure}

% =============================================================== %
\section{Results and Discussion\label{sec:results}}
\subsection{Pareto Dominant Student MLIPs \label{sec:pareto}}
Atomistic simulations require accurate and efficient evaluation of energy and forces. Performing MD at large length and time scales is challenging due to the trade-off between accuracy and efficiency of the interatomic potential used in the simulation. MLIPs provide a path to perform MD with QC accuracy at dramatically reduced costs relative to methods that use explicit QC solutions to calculate the forces and energies that are used to evolve the simulation forward in time. 
Here, we show that the teacher-student method allows us to improve the accuracy of the smaller MLIPs without raising the computational costs at inference time. 

Characterizing the accuracy of interatomic potentials in absolute terms is challenging~\cite{liu2023discrepancies}. Such comparisons should be made against exact numerical solutions or experimental data, which may not be available. The accuracy of MLIPs can be judged against held-out test data from the QC calculations used to develop potential. In this section, we use the force root mean squared error (RMSE) as an accuracy measure. We analyze out-of-sample MD-based accuracy metrics in Sec~\ref{sec:accuracy}. 

It is also difficult to define an unambiguous metric of efficiency for MLIPs. An informative measure is the time-per-MD step per atom, which is affected by hardware (CPU/GPU memory and processing speeds) and software (MD library, algorithm used for neighbor list construction, etc.). Therefore, in this article, we use a consistent hardware configuration (a single A6000 GPU with 48 GB of storage) and we use the Atomic Simulation Environment~\cite{larsen2017atomic} software. We simulate $48,000$ atoms at fixed volume and energy (NVE ensemble) with a time step of 1\,fs and a simulation time of 1\,ps. Additionally, we can characterize the efficiency of MLIPs based on memory requirements. We record the maximum number of atoms that can be simulated on a single A6000 GPU using the ASE software and use it to calculate the average memory-per-atom. 

Figure~\ref{fig_pareto} shows the Pareto plot of accuracy (force RMSE) against the time per MD step per atom and maximum atoms per GPU in panels (A) and (B) respectively. The student models are Pareto dominant with respect to the control MLIPs, i.e., for a given cost (MD speed or memory requirement) the student models have higher accuracy than the control models. Each data point is averaged over four MLIPs initialized with different random seeds. Relative to the teacher, the student MLIPs can simultaneously achieve more than a factor of 2 speed up in MD simulations at less than half the memory requirements, while sacrificing less than 20\% in force accuracy.

% ***************************************** %
\begin{figure*}
    \includegraphics[]{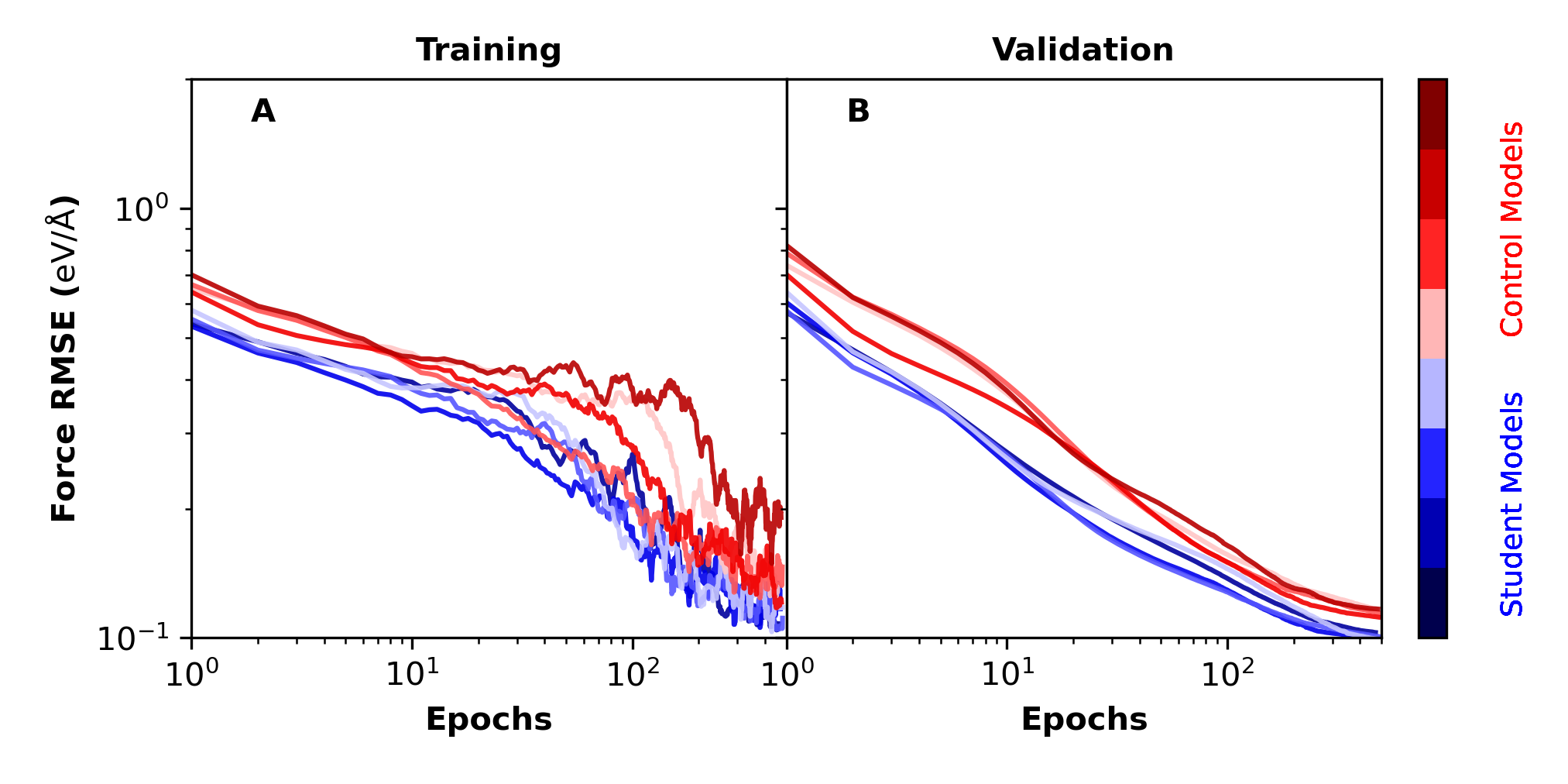}
\caption{\textit{The student HIPNN MLIP exhibits faster learning dynamics than the control models.} 
Plot of force RMSE errors for the training and validation data splits are shown as a function of epochs for student and control MLIPs in (A) and (B) respectively. 
}
\label{fig_learning-dynamics}
\end{figure*}  
% ***************************************** %

% ================================================================= %
\subsection{Accuracy and MLIP Capacity\label{sec:accuracy}}
In Fig.~\ref{fig_accuracy}, we analyze several metrics of accuracy such as force RMSE errors, energy-per-atom RMSE errors, and radial distribution function (RDF)~\cite{allen2017computer} error as a function of the number of weights of each HIPNN model, which serves as an effective measure of the model capacity. This metric is compelling because it is agnostic to the hardware constraints, such as GPU memory and the choice of MD software. While it is meaningful to compare weights for different variants of a MLIP architecture (HIPNN), care should be taken when making comparisons across different architectures. In Sec.~\ref{sec:speed-scale}, we study how the number of weights correlates with the maximum number of atoms per GPU and MD speeds. 

The force RMSE errors in panel (A) of Fig~\ref{fig_accuracy} indicate a meaningful improvement of the student models compared with the control models, while the energy errors in panel (B) are only minimally improved in the student models. 
We compute the RDFs of liquid aluminum at 1200\,K using the LAVA~\cite{dang2023lava} software, which is a wrapper for LAMMPS~\cite{thompson2022lammps}. The reported RDF errors shown in panel (C) for the student and control models are the total absolute errors with respect to the teacher RDFs. 
Note that the energy and forces errors in Fig.~\ref{fig_accuracy}(A) and (B) respectively are computed with respect to the ground truth training data, whereas the RDF errors in (C) are the differences against the teacher models' MD simulations. Figure~\ref{fig_accuracy} reveals that the energy RMSE is quite similar between the student and control models, except for very small model sizes. We attribute the improvements in the force accuracy to the fact that the atomic energies are local properties predicted for each atom, similar to atomic forces. Furthermore, the force errors strongly correlate with the RDF errors. Note that the RDF simulations constitute a difficult extensibility test, because the simulations utilize large periodic boxes that are two orders of magnitude larger than the configurations found in the training data, with each simulation containing 18634 atoms.

% =========================================================================================== %
\subsection{Learning Dynamics~\label{sec:learning-dynamics}}
Learning dynamics (also known as optimization dynamics) characterize how the training or out-of-sample validation error evolves as a function of the training epochs. We analyze learning dynamics of the student and control MLIPs to understand the how the auxiliary targets affect the training. We see in Fig.~\ref{fig_learning-dynamics} that the student models' learning dynamics out perform the control models. Loss curves of four different student and control models, which all share the same architecture. Here, we use four different seed values for the random initialization of the sets of four control and four students respective. By employing the same seed for a pair of student and control MLIPs, we ensure that each student MLIP has the same initial weights as its corresponding control MLIP. Note that the learning dynamics curves begin after one epoch of training has been done. The apparent improved initialization of the student models are actually due to the improved learning by the student HIPNN models due to the atomic energy targets from the teacher. We also note that the auxiliary targets have a regularization effect, stabilizing the optimization, because the learning curves of the student models exhibit less variance in the later stages of training.

% ================================================================ %
\subsection{Speed and Scalability~\label{sec:speed-scale}}
Performing large-scale MD is necessary to address many important scientific questions. While the atomic time-steps are generally of order 1--10\,fs, the physical phenomena of interest may span milliseconds or longer and can require billions of atoms. Large-scale MD on modern super-computing clusters relies on the idea of weak scaling~\cite{thompson2022lammps}, where the simulation is distributed across many nodes. Due to the high latency of inter-node communications, it is beneficial to fit as many atoms on one node or GPU as possible. 

\begin{figure}[]
    \includegraphics[]{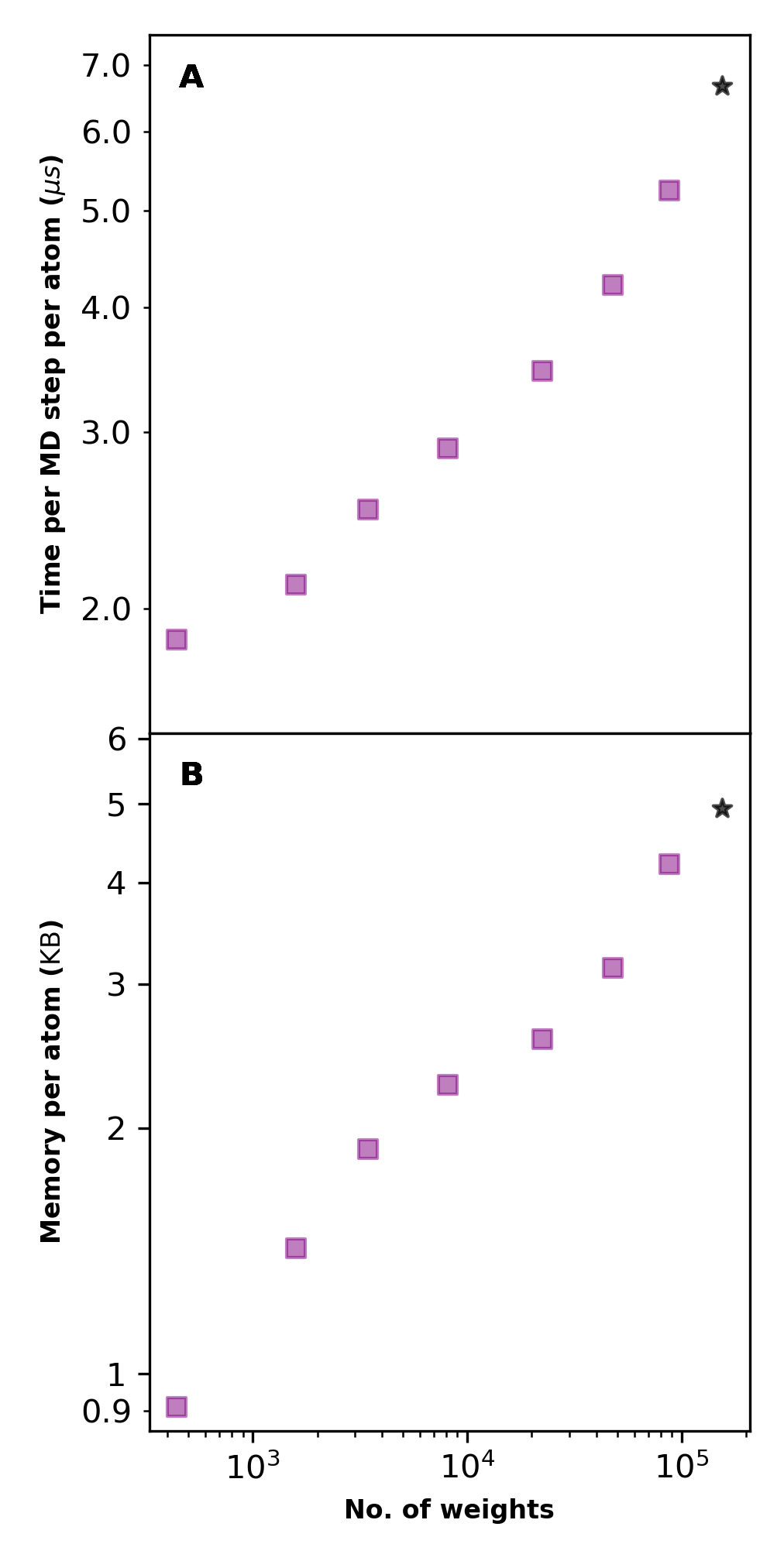}
\caption{\textit{Inference efficiency as measured by (A) time-per-MD step per atom and (B) memory per atom, versus the number of trainable weights in HIPNN models}. Data is collected by running MD simulations using ASE in the NVE ensemble on a single A6000 GPU (48GB) for $10^3$ steps. 
}
\label{fig_scaling}
\end{figure}

Our work shows that the teacher-student procedure lowers the resource required for large-scale MD at a target accuracy. The teacher-student method improves the accuracy of smaller models. The light-weight student models require fewer floating point operations for force evaluations. As a direct consequence, we can now run large scale MD at a desired accuracy while using fewer computational resources.
Fig.~\ref{fig_scaling} examines how the number of weights in the network affect the computational cost for large-scale MD in terms of both speed and memory in  panels (A) and (B) respectively. 
The MD is performed using ASE in the NVE ensemble for 1\,ps with a time step of 1\,fs.  Furthermore, we see that the smaller student models can fit more atoms on a single GPU (A6000 48 GB in Fig.~\ref{fig_scaling}). The ability to fit more atoms on a single GPU can improve the efficiency of large MD simulations, because inter-node latency can often dominate the computations~\cite{thompson2022lammps}. 

% ================================================================ %
\subsection{Student MLIP Surpasses the Teacher~\label{sec:Born-Again}}
We apply the ``Born Again''~\cite{furlanello2018born} teacher-student training to HIPNN. The `Born Again' method was introduced in the Ref~\cite{furlanello2018born}, and it is a variant of the teacher-student training where both models have exact same architecture and number of trainable weights. The central aim of the `Born Again' method is to train a student model that surpasses the teacher's accuracy. This is in contrast to knowledge distillation, where the aim is to improve the accuracy of smaller student models that have fewer trainable weights than the teacher models.
We show that student models trained to the ground truth data and the teacher model's auxiliary outputs and can surpass the teacher by using an loss scheduler. 

Initially, we use the static loss function in Eq.~\ref{eq_loss} to find that student MLIPs achieve comparable errors to the teachers. However, then we utilize a loss scheduler to dynamically update the weights $w_A, w_F, w_E$ during the training as summarized in Table~\ref{tabel:loss_scheduler} in Appendix~\ref{app:hyper}. The weights in the loss function were chosen to favor the teacher's knowledge in the early stages of the training and emphasizing the QM data in the later stages, so that the student can surpass the accuracy on the QM. The values of the weight schedule were determined through manual hyper parameter tuning. The final weights of the students' loss function match that of that of the static weights of the teacher MLIP. The student models' Energy RMSE of $0.37 \pm 0.02~\mathrm{eV}$, and force RMSE of $0.083 \pm 0.003~\mathrm{eV/\si{\angstrom}}$ is lower than the teacher models errors, energy RMSE of $0.38 \mathrm{eV}$, and force RMSE of $0.092~\mathrm{eV/\si{\angstrom}}$. This shows that the BA teacher-student approach improves the force accuracy by about $10\%$, which shares the same model architecture and underlying training data.

% ================================================================ %
\section{Conclusions and Future work\label{sec:conclusion}}
We introduce a teacher-student framework that can be readily applied to many MLIPs that decompose the configurational energy into a sum of local contributions. The teacher-student training framework improves the Pareto set of error-cost trade-offs for MLIPs, yielding models that are computationally cheaper for the same accuracy,  or more accurate for the same computational cost, and requires no additional ground-truth data.

In a practical setting, we showed that the student MLIPs are Pareto dominant with respect to the control MLIPs. The student MLIPs, trained to the ground truth and teacher's atomic energies, achieve higher accuracy at the same efficiency (speed and memory requirements) when compared to the control models that were only trained to ground truth.
We use both training errors (energy and force RMSEs) as well as MD based metric (RDF errors) to quantify the accuracy of the MLIPs. We find that the force RMSE errors generally correlate with the MD based metrics, which are costlier to evaluate. MD based metrics also demonstrate the extensibility of the MLIP to configurations much larger than what is typically included in the training dataset.
However, MD based metrics require a ground truth, such as ab initio results or experimental data, which may not always be available. Efficiency of MLIPs is also multi-faceted. 
We use MD speeds and memory-per-atom as effective measures. While MD speed is easy to interpret, it strongly depends on the underlying hardware, MD codes, and the system under study.
The memory-per-atom measure is useful to optimize for large MD simulations, where inter-node communications may dominate. While the research to improve the accuracy of MLIPs has focused on the development of more expressive architectures~\cite{duval2023hitchhiker}, and the generation of larger datasets~\cite{kaplan2025foundational}, we show that the `born-again' inspired teacher-student training can be used to train more accurate models with existing datasets and architectures. Thus, innovations to the training protocol can allow us to extract better models from existing datasets. 

We used the atomic energies as the knowledge for the teacher-student-training. 
In the language of message passing graph neural networks, the atom energy is a node level scalar~\cite{duval2023hitchhiker}. In future works, one can explore node level vectors, such as forces~\cite{gong2025predictive, matin2025ensemble}, and edge level properties as knowledge for the teacher-student framework to train across different architectures~\cite{kelvinius2023accelerating, amin2025towards}.
Additionally, network weights for the interactions layers may be utilized as knowledge for the student. One may also explore using an ensemble of teachers for future work, where the students may be able to leverage the uncertainty associated with the teacher MLIPs' knowledge. 

Graph neural networks can be used to predict molecular properties beyond just energy and forces such as charges~\cite{nebgen2018transferable, ko2021fourth}, dipoles~\cite{sifain2018discovering}, and other properties~\cite{fedik2022extending}. There is potential to explore cross-modal teacher-student training frameworks to combine teacher models with different specialized tasks to train generalist student MLIPs. This will be an important step towards foundation models for chemistry with broad applicability.

\appendix
\section{Training Details}
\subsection{Hyper-Parameters\label{app:hyper}}
We use HIPNN models with $1$ interaction layer. The teacher models use $4$ atom layers (feed-forward layers) with a width of $128$. The student (and control) models have between $1$ and $4$ atom layers with width between $12$ and $128$. All models have maximum tensor sensitivity order set at $l_\mathrm{max}=1$. For the sensitivity functions which parameterize the interaction layer, radial basis functions are used with a soft-min cutoff of $1.5~\si{\angstrom}$, the soft maximum cutoff of $7.0~\si{\angstrom}$, and hard maximum cutoff of $7.5~\si{\angstrom}$. The teacher model uses $40$ basis functions. For the student (and control) MLIPs, we use between $8$ and $40$ sensitivity functions. The soft-min cut-off corresponds to the inner cut-off at very short distances. The hard maximum cut-off corresponds to the long distance cut-off. The soft maximum cutoff is set to a value smaller than the hard-dist cutoff to ensure a smooth truncation of the sensitivity functions. Note that the we are using the naming conventions for the hyper-parameters in the HIPNN GitHub Repository~\cite{lanl2021hippynn}, which differs slightly from the original HIPNN publication~\cite{lubbers2018hierarchical}. 

We summarize the weights corresponding the loss function in Eq.~\ref{eq_loss}. $W_{L_2}=10^{-6}$ and $W_{R}=0.01$ is common to the teacher, student and control models. $W_E=1$ and $W_F=10$ for the teacher and control MLIPs. Lastly, we used $W_E=1$, $W_F=30$ and $W_A=100$ for the student models. For the BA training, we use a loss schedule with weights given in Table~\ref{tabel:loss_scheduler}.

We used the Adam Optimizer, with an initial learning rate of $0.001$, which is halved with a patience of $30$ epochs. The termination patience is $50$ epochs.

\begin{table}[h]
\small
  \caption{\ Loss Scheduler for Student MLIPs for the Born-Again method of teacher-student training in Sec.~\ref{sec:Born-Again} }
  \label{tabel:loss_scheduler}
  \begin{tabular*}{0.48\textwidth}{@{\extracolsep{\fill}}llll}
    \hline
    Epoch  & $w_A$ & $w_F$ & $w_E$ \\
    \hline
    1   & 200 & 75 & 0.0  \\
    200 & 160 & 63 & 0.2 \\
    250 & 120 & 51 & 0.4 \\
    300 & 80 & 39 & 0.6 \\
    350 & 40 & 27 & 0.8 \\
    400 & 0 & 15 & 1 \\
    \hline
  \end{tabular*}
\end{table}

\section*{Conflicts of interest}
There are no conflicts to declare.

\section*{Data availability}
The HIPNN~\cite{lubbers2018hierarchical, chigaev2023lightweight} MLIP is implemented in an open-source PyTorch-based software package called \texttt{hippynn} which is available for download~\cite{lanl2021hippynn}. We will upload the scripts for workflow and trained models to the examples folder. We use an aluminum dataset published in Ref.~\cite{smith2021automated} and the dataset is available for download~\cite{atomistic-ml2021ani-al}.

\section{Author contributions}
Conceptualization (SM, NL); Methodology (SM, AEAA, ES, AP, GTC, BN, JSS, RM, YWL, ST, KB, NL); Formal analysis (SM); Writing - original draft (SM); Writing - review and editing (SM, AEAA, ES, AP, GTC, BN, JSS, RM, YWL, ST, KB, NL); Supervision (GTC, NL).

\section*{Acknowledgements}
This work was supported by the United States Department of Energy (US DOE), Office of Science, Basic Energy Sciences, Chemical Sciences, Geosciences, and Biosciences Division under Triad National Security, LLC (‘Triad’) contract grant no. 89233218CNA000001 (FWP: LANLE3F2, LANLE8AN).
We acknowledge support from the Los Alamos National Laboratory (LANL) 
Directed Research and Development funds (LDRD).
This research was performed in part at the Center for Nonlinear Studies (CNLS) at LANL. 
This research used resources provided by the Los Alamos National Laboratory Institutional Computing Program and Darwin testbed at LANL which is funded by the Computational Systems and Software Environments subprogram of LANL's Advanced Simulation and Computing program.

\bibliography{References} 
\bibliographystyle{rsc} %the RSC's .bst file

% ===============================================================%
\begin{figure*}
    \centering
    \includegraphics[width=0.75\textwidth]{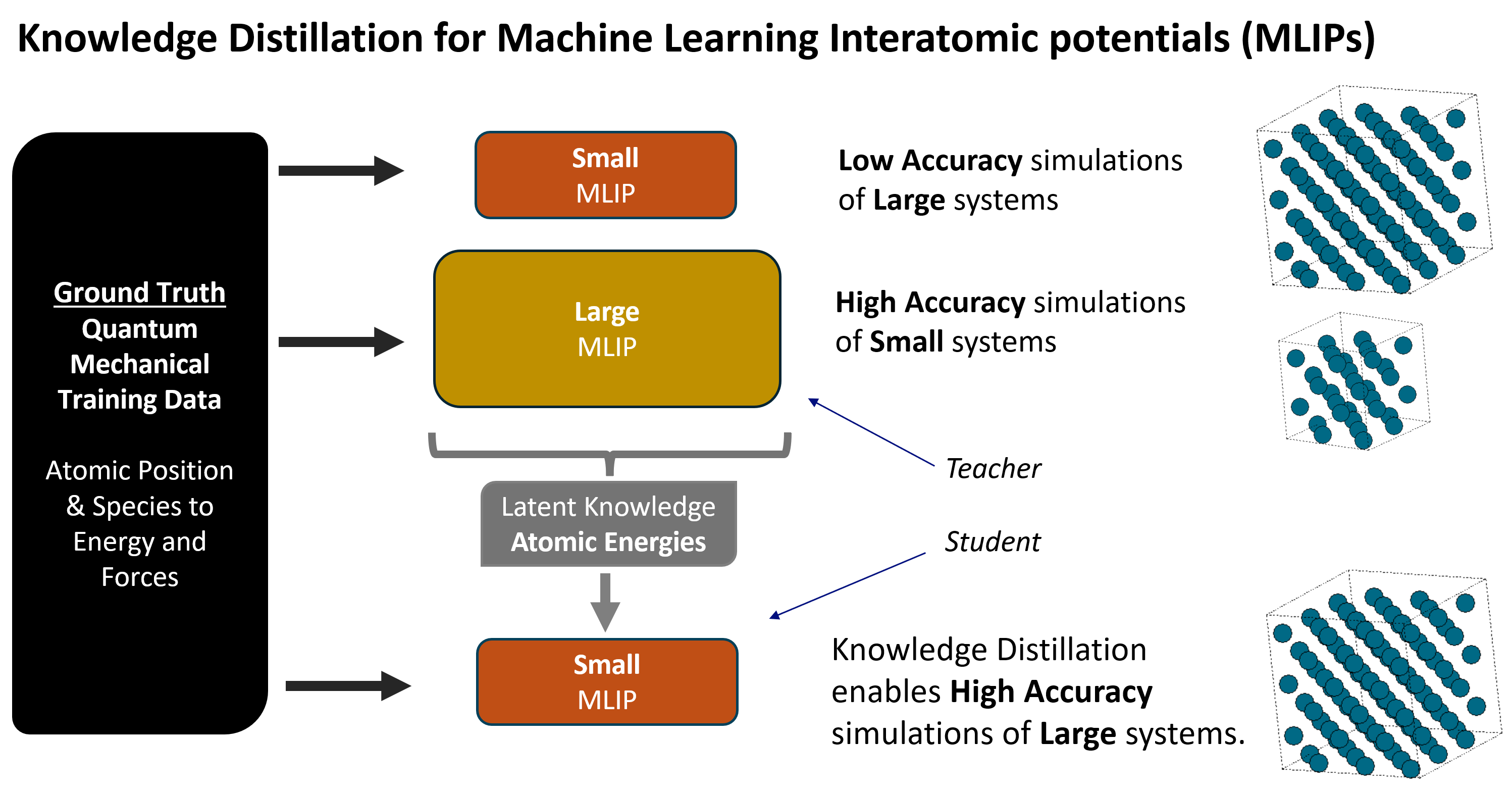}
    \label{fig:TOC}
\end{figure*}

\end{document}